\documentclass[lnbip]{svmultln}

\pdfoutput=1

\usepackage{makeidx}
\usepackage{amssymb}
\usepackage{graphicx}
\usepackage{url}

\urldef{\mailsa}\url{adrian.paschke@inf.fu.berlin.de}
\urldef{\mailsb}\url{schaef@inf.fu-berlin.de}

\begin{document}

\mainmatter  

\title{Aspect OntoMaven - Aspect-Oriented Ontology Development and Configuration With OntoMaven}

\titlerunning{Aspect OntoMaven}

\author{Adrian Paschke \and Ralph Schaefermeier}
\authorrunning{Adrian Paschke and Ralph Schaefermeier}

\institute{Corporate Semantic Web, Institute of Computer Science, FU Berlin\\
\mailsa\\
\mailsb
}

\toctitle{Aspect OntoMaven - Aspect-Oriented Ontology Development and Configuration With OntoMaven}
\tocauthor{Aspect-Oriented Ontology Development and Configuration}
\maketitle

\begin{abstract}
In agile ontology-based software engineering projects support for modular reuse of ontologies from large existing remote repositories, ontology project life cycle management, and transitive dependency management are important needs. The contribution of this paper is a new design artifact called OntoMaven combined with a unified approach to ontology modularization, aspect-oriented ontology development, which was inspired by aspect-oriented programming. OntoMaven adopts the Apache Maven-based development methodology and adapts its concepts to knowledge engineering for Maven-based ontology development and management of ontology artifacts in distributed ontology repositories. The combination with aspect-oriented ontology development allows for fine-grained, declarative configuration of ontology modules.
\end{abstract}

\section{Introduction}
Sharing and reusing knowledge in ontology-based applications is one of the main aims in the Semantic Web as well as the Pragmatic Web\footnote{\scriptsize{\url{http://www.pragmaticweb.info}}} \cite{DBLP:conf/ruleml/WeigandP12,DBLP:journals/ijait/PaschkeB11,DBLP:conf/icpw/PaschkeBKC07}, which requires the support of distributed ontology management, documentation, validation and testing.  Typically ontologies are developed and maintained in an iterative and distributed way, which requires the support of versioning \cite{conf/gi/Luczak-RoschCPRT10,PA155} and modularization \cite{CLHP2009a,Coskun2012}.
Aspect-Oriented Ontology Development (AOOD) \cite{SchPa13, schafermeier:2014aa} enables weaving of  cross-cutting knowledge concerns into the main ontology model, which requires meta-level descriptions of ontology aspects and management of distributed knowledge models.

In this paper we introduce OntoMaven\footnote{\scriptsize{\url{http://www.corporate-semantic-web.de/ontomaven.html}}}, which adapts a highly successful method and tool in distributed software engineering, namely Apache Maven\footnote{\scriptsize{\url{http://maven.apache.org/}}}, for the Maven-based management of distributed ontology repositories. The OntoMaven approach supports ontology engineering in the following ways:

\begin{itemize}
\item OntoMaven remote repositories enable distributed publication of ontologies as \textbf{ontology development artifacts} on the Web, including their metadata information about life cycle management, versioning, authorship, provenance, licensing, knowledge aspects, dependencies, etc.
\item OntoMaven local repositories enable the reuse of existing ontology artifacts in the users' local ontology development projects.
\item OntoMaven's support for the different development phases from the design, development to testing, deployment and maintenance provides a flexible life cycle management enabling iterative agile ontology development methods, such as COLM \footnote{\scriptsize{\url{http://www.corporate-semantic-web.de/colm.html} \cite{RoeHe09}}}, with support for collaborative development by, e.g., OntoMaven's dependency management, version management, documentation and testing functionalities, etc.
\item The extension of OntoMaven with aspect-oriented concepts allows the declarative configuration and automated interweaving of ontology modules during the build phase.
\item OntoMaven plug-ins provide a flexible and light-weight way to extended the OntoMaven tool with existing functionalities and tools, such as semantic version management (e.g., SVont - Subversion for Ontologies \cite{conf/gi/Luczak-RoschCPRT10,PA155}), semantic documentation (e.g., SpecGen Concept Grouping \cite{Coskun2012}), dependency management of aspect-oriented ontology artifacts (e.g. \cite{SchPa13}), automated testing (e.g., with the W3C OWL test cases and external reasoners such as Pellet), etc.
\item Maven's API allows easy integration of OntoMaven into other ontology engineering tools and their integrated development environments (IDE).
\end{itemize}

The further paper is structured as follows: Section \ref{OntoMavenConcept} describes the conceptual design of OntoMaven based on Maven's remote and local repositories, the Project Object Model (POM), and plug-ins. Section \ref{AOOD} introduces the approach to aspect-oriented ontology development. Section \ref{proof-of-concept} proves the feasibility of the proposed concepts with a proof-of-concept implementation of the OntoMaven design artifact. Section \ref{relatedwork} compares the OntoMaven functionalities to the tool support of the major existing ontology engineering tools. Finally, section \ref{conclusion} summarizes the current OntoMaven work and discusses future research.

\section{OntoMaven's Design and Concept }
\label{OntoMavenConcept}

In the following subsections we adapt the main concepts of Maven, so that they can be used in ontology development and ontology life cycle management. In particular, we focus on the (distributed) management of knowledge artifacts (ontologies / ontology modules) and their versioning, import and dependency management, module configuration, documentation, and testing.

\subsection{Management and Versioning of Ontology Artifacts}
\label{versioning}

Typically, in ontology reuse there is a need for versioning and life cycle management. Combinations with existing ontologies (by ontology matchmaking and alignment) might lead to transitive dependencies which need to be described and managed. OntoMaven therefore adopts Maven's artifact concept. It describes and manages ontologies as ontology artifacts in a Maven Project Object Model (POM). The typical steps to add an ontology (module) as an OntoMaven artifact to a POM are:

\begin{enumerate}
  \item Find ontology module(s)
  \item Select the right module and version
  \item Analyse and resolve dependencies of the modules
  \item Declaratively describe the ontology artifact in a POM
\end{enumerate}

Many ontology languages support imports or integration of distributed ontologies. For instance, the W3C Web Ontology Language (OWL) therefore has a specialized \texttt{owl:import} statement. Typical recurring tasks which are automated by OntoMaven in such modular import and reuse scenarios are,

\begin{itemize}
  \item check the existence of the imported ontology (module) referenced by the defined URI in the import statement (and find alternative URLs from pre-configured repositories if the ontology does is not found at the import URI).
  \item management of ontologies / ontology modules as ontology artifacts in Maven repositories including their metadata descriptions such as versioning information.
  \item download of ontology artifacts from remote repositories (including transitive imports) to a local development repository in order to support offline development of ontologies
\end{itemize}

Another important aspect in the agile and collaborative development of ontologies is the support for version management. Typical requirements are maintaining consistency and integrity, as well as provenance data management (e.g. authorship information) throughout the version history. The approach in OntoMaven is based on the ontology versioning tool \emph{SVont}, which is an extension of the version management tool Subversion. \cite{conf/gi/Luczak-RoschCPRT10,PA155}

\subsection{Documentation}
Documentations of ontologies ease maintenance and reuse. The typical distinction is into \emph{user documentation} and \emph{technical documentation}. While the former supports the users of an ontology, e.g. in their task to populate the ontology with instance data, the latter, technical documentation, supports the ontology developer. Ontology concept groupings and summarizations of concepts provides the documentation reader with an easier way to understand the ontology vocabulary. \cite{Coskun2012} Maven supports the documentation phase and provides goals for creating and publishing automated reports. In OntoMave we make use the SpecGen extension\footnote{\scriptsize{\url{http://www.corporate-semantic-web.de/concept-grouping.html}}} for automated concept grouping, in order to create the technical and user documentation in an OntoMaven plugin which is executed by the \texttt{mvn site} command.

\subsection{Testing}
Testing is an important phase in the ontology life cycle. In particular, in agile iterative development processes testing allows detecting inconsistencies, anomalies, improper design, as well as validation against, e.g., the intended results of domain experts' competency questions which are represented as ontology test cases. Maven supports a testing phase in which automated tests are executed and the results are reported by the Maven command \texttt{mvn test}. As standard test types OntoMaven by default supports the W3C OWL test cases\footnote{\scriptsize{\url{http://www.w3.org/TR/owl-test/}}} \emph{syntax checker}, \emph{consistency checker}, and \emph{entailment test}. The produced test results are compliant to the W3C recommendation and the created test reports show if the ontology model is \texttt{consistent}, \texttt{inconsistent} or if the result is \texttt{unknown}. Further test types can be implemented as Maven plug-ins and added to the OntoMaven projects test suites by the user.

\section{Aspect-Oriented Ontologies}
\label{AOOD}

Aspect-oriented design and programming are paradigms in software development and are means for system modularization by separation of \emph{cross-cutting concerns}.
In software systems, typical cross-cutting concerns comprise logging, authentication, and transaction management.
In \cite{schafermeier:2014aa}, we proposed the idea of aspect-oriented ontologies, providing an approach to ontology modularization based on cross-cutting concerns.

As in software development, cross-cutting concerns are linked to requirements.
Requirements can be functional, i.e., directly related to the business goals the system or ontology is supposed to accomplish.
Non-functional requirements, in contrast, are related to goals concerning the system or ontology itself.
Functional ontology requirements are generally directly related to the competency questions the ontology is supposed to answer, while examples for non-functional requirements include provenance information, multilinguality and reasoning complexity.

Each of these requirements is mapped to one ore multiple sets of axioms in an ontology. If there exists a relationship between an aspect and an axiom, it means that this axioms belongs to the aspect in question. This way, an ontology can be modularized into (possibly overlapping) modules, each module representing a requirement, and each module-requirement pair being encapsulated in an aspect.

Aspects can, in turn, be formal descriptions themselves, i.e., just as the ontology module they are mapped to, they can be ontological statements as well. For example, a set of facts in an OWL ontology can be related to a temporal aspect (e.g. \emph{Bonn is capital of West Germany}, which was only valid from 1949 to 1990), where the temporal aspect is an OWL individual that comes with relations and properties formalized using the W3C time ontology and representing the time interval \emph{1949-1990}.

The individual representing the aspect may be directly referenced by its IRI (if it is a named individual). Beyond that, it is possible to define super aspects by using some sort of query (e.g. \emph{all the things that happened during the 20th century} which is equivalent to all facts in the ontology that are related to temporal aspects which are valid between 1900 and 1999).

\begin{figure}[hbt]
  \centering
  \includegraphics[width=.8\textwidth]{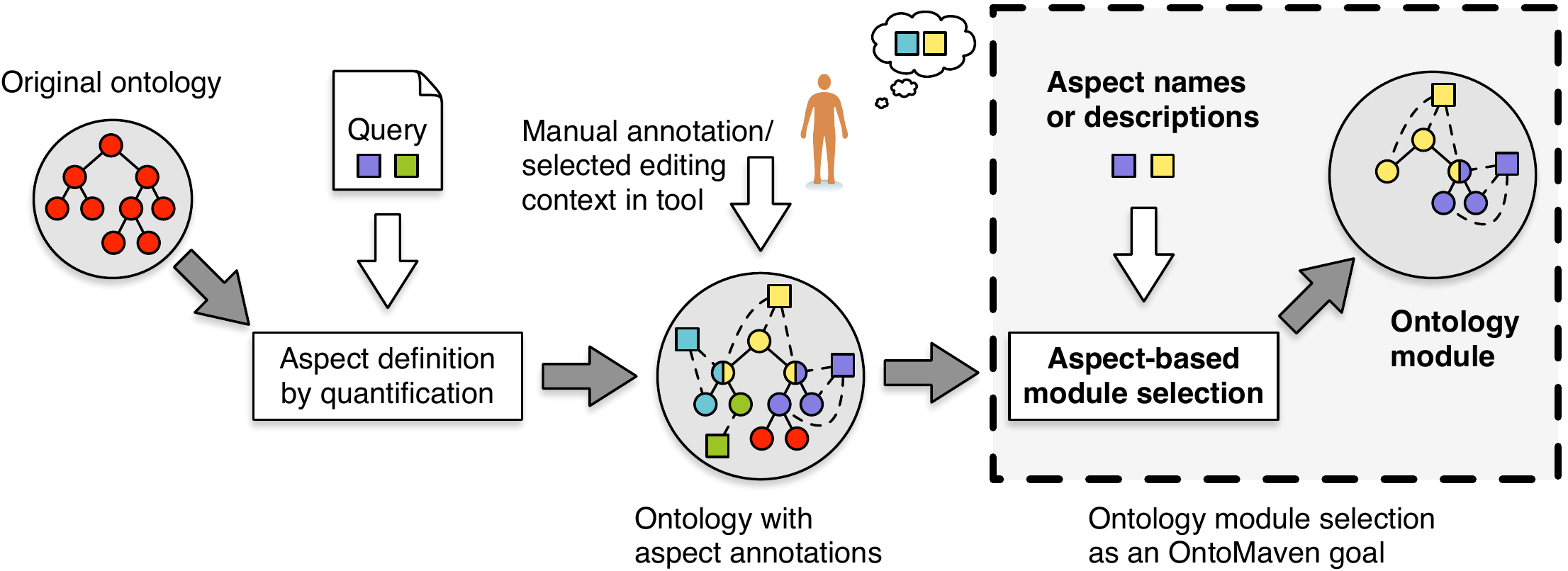}
  \caption{Depiction of the ontology aspect configuration plugin for OntoMaven (dashed box on the right). One or several aspect names are passed to the OntoMaven plugin as arguments. The result is an ontology module including exactly those axioms that belong to the given aspects. Axioms of the original ontology (left) have been annotated with entities from an external aspect ontology (center). Axiom declaration is based on queries or is performed manually.}
  \label{fig:aood_approach}
\end{figure}

The left part of Figure \ref{fig:aood_approach} depicts the process of enriching an ontology with aspects.

When the ontology is deployed in the context of an application, modules can be recombined by directly referencing their URIs or by using queries as described above. This may happen either dynamically at runtime or during the configuration phase of the application in a descriptive manner.

In the following section, we describe how aspect composition has been implemented as a configuration step in the application and ontology management lifecycle of OntoMaven.

\section{Proof-of-Concept Implementation - OntoMaven PlugIns}
\label{proof-of-concept}

The implementation of OntoMaven \cite{kilic13} extends and adapts Maven, so that it supports the management of ontology modules in Maven repositories. This section describes how the OntoMaven approach, using \emph{Maven repositories} and the \emph{Maven plug-in} extension mechanism. A Maven plug-in is a collection of one or more goals.
The implementation of a Maven plug-in is done in an \emph{Maven Plane Old Java Object (MOJO)}. In the OntoMaven approach, the phases and goals, which the plug-in implements, are defined by JavaDoc annotations in the source code of the Mojo class.
For instance, the following annotations define that the implemented plug-in is used in the phase \texttt{test} and that is has a goal called \texttt{test-syntax}:

\begin{scriptsize}
\begin{verbatim}
@phase test //plug-in used in test phase
@goal test-syntax // goal with the name "test-syntax"
\end{verbatim}
\end{scriptsize}

Parameters are used to configure the plug-in execution. For instance, the following code snippet defines a required parameter \texttt{compliancemode} with the default value \texttt{strict}:

\begin{scriptsize}
\begin{verbatim}
*@parameter expression = “compliancemode“ @default-value="strict" @required
\end{verbatim}
\end{scriptsize}

Such plug-in parameters can be configured in a POM.xml file or directly when calling a goal, e.g. \texttt{mvn ... -Dcompliancemode=strict}.
An implemented plug-in can be installed using Maven \texttt{mvn install} and the plug-in goals can be integrated into the POM.xml of an OntoMaven project, as the following example listing shows for the plug-in SVontPlugin and the goal semantic-diff:

\begin{scriptsize}
\begin{verbatim}
<build>
    <plugins> <plugin>
            <groupId>de.csw.ontomaven</groupId>
            <artifactId>SVontPlugin</artifactId>
            <version>1.0-SNAPSHOT </version>
            <executions> <execution>
               <goals> <goal>semantic-diff</goal>
...
\end{verbatim}
\end{scriptsize}

The following subsections provide further details about the proof-of-concept implementations of the main plug-ins in OntoMaven. We first describe the OntoMaven repositories which are the persistence and back-end layer for storing and managing ontologies.

\subsection{OntoMaven Repositories}
\begin{figure}
  \centering
  \includegraphics[width=6cm]{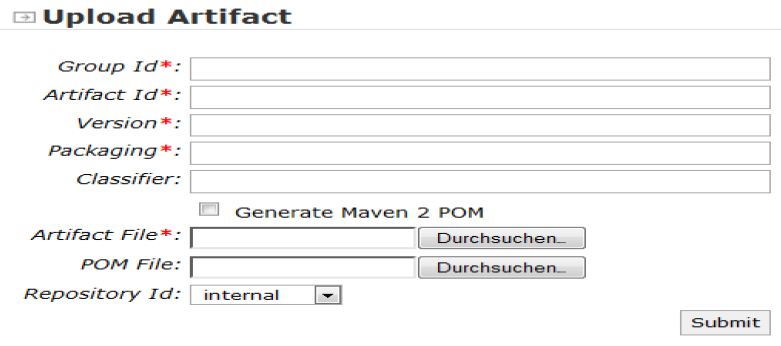}
  \includegraphics[width=6cm]{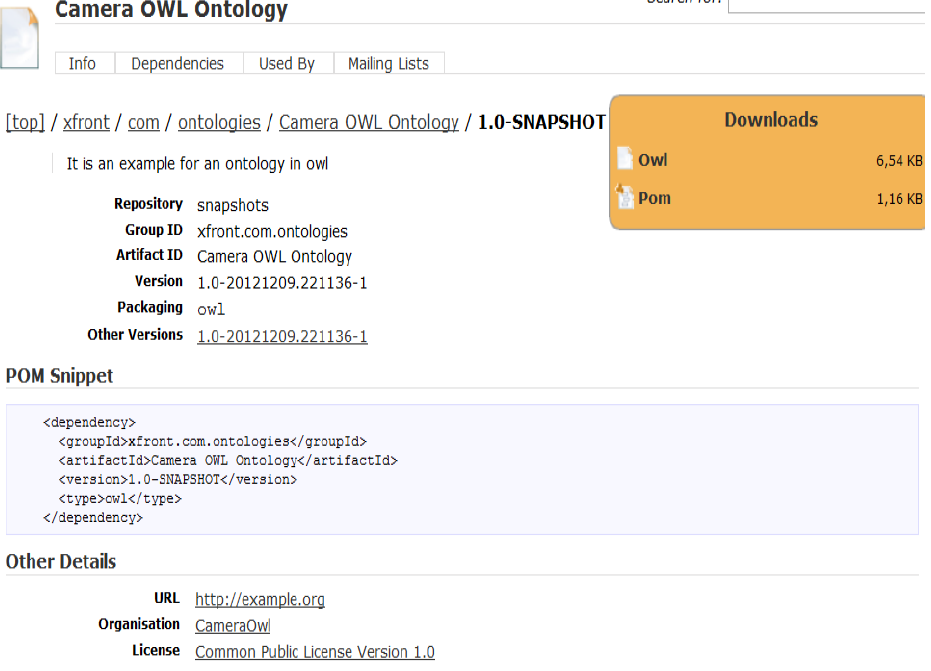}
  \caption{Archiva User Interface - Ontology Artifact Upload and Management}\label{fig:ArchivaUpload}
\end{figure}

OntoMaven can use all Maven compliant repositories. One of the strengths of Maven is that is uses a folder structure following a standard folder layout for its repositories. For the OntoMaven proof-of-concept implementation we adapted the Apache Archiva Build Artifact Repository Manager\footnote{\scriptsize{\url{http://archiva.apache.org/}}} as a managing tool providing a user interface for the OntoMaven repositories. The left figure \ref{fig:ArchivaUpload} shows the upload user interface.

Via this form an ontology can be uploaded to an OntoMaven repository together with its POM file. The artifact's metadata contains information about the group id, artifact id, version, packaging and optional additional classifier information. The POM provides all necessary information about the artifact and its dependencies. In OntoMaven these dependencies are used to describe (transitive) imports from an ontology, which are resolved by the OntoMvnImport plug-in (see subsection \ref{OntoMvnImportPlugIn}).

The right figure \ref{fig:ArchivaUpload} gives an example of the Archiva user interface showing the management information of an ontology artifact called \texttt{Camera OWL Ontology}. Under the interface menu link \texttt{Dependencies} the dependencies of this ontology can be found.

Once managed in an online OntoMaven repository, an ontology artifact can be used in any OntoMaven ontology development project. The following listing gives an example how a remote repository can be configured and a dependency to an ontology artifact (here the Camera-OWL-Ontology) can be defined in the POM.xml document of a project.

\begin{scriptsize}
\begin{verbatim}
<profiles> <profile>
        <id>2</id>
        <activation> <activeByDefault>true</activeByDefault> </activation>
        <repositories> <repository>
                <snapshots> <enabled>true</enabled> </snapshots>
                <id>snapshots</id>
                <name>OntoMaven Snapshot Repository</name>
                <url>http://www.corporate-semantic-web.de/repository/snapshots/</url>
    ...
</profiles>

<dependencies> <dependency>
        <groupId>xfront.com.owl.ontologies</groupId>
        <artifactId>Camera-OWL-Ontology</artifactId>
        <version>1.0-SNAPSHOT</version> <type>owl</type>
</dependency> </dependencies>
\end{verbatim}
\end{scriptsize}

\subsection{OntoMvnImport}
\label{OntoMvnImportPlugIn}

This plug-in implements the imports of ontologies into the Maven repositories. It  also checks if the import statements in the ontology (including transitive imports) can be resolved. Therefore, it maintains an updated list of referenced URIs to the ontology resources loaded to the repository.
This list follows the OASIS XMLCatalog standard which also specifies a technique for the automated replacement of external references in XML documents. In OntoMaven we use this automated replacement technology to replace the URI references to imported \textbf{external ontologies} with references to the \textbf{internal ontology artifacts}, which are locally managed in an OntoMaven repository. This replacement approach avoids the continuous import and use of external ontologies during an OntoMaven development project. After the first loading of an ontology as repository artifact by the OntoMvnImport plug-in, the plug-in always checks if there is an ontology artifact listed in the XMLCatalog. If there is an existing reference to an ontology artifact, it will use it instead of any externally referenced ontology. The following listing shows how the plug-in can be used in a OntoMaven POM. In the \texttt{configuration} it defines the input ontology and sets the \texttt{local} parameter to true, indicating that the ontology should be loaded to the local repository and that the local version of the ontology should be used.

\begin{scriptsize}
\begin{verbatim}
<build> <plugins> <plugin>
    <groupId>de.csw.ontomaven</groupId>
    <artifactId>OntoMvnImport</artifactId>
    <version>1.0-SNAPSHOT</version>
    <configuration>
        <owlfile>src/resource/reputation.owl</owlfile> <local>true</local>
    </configuration>
    <executions> <execution>
        <goals> <goal>owlimport</goal> </goals>
...
\end{verbatim}
\end{scriptsize}

\subsection{OntoMvnApplyAspects}
\label{OntoMvnApplyAspects}

This plug-in contributes to the \texttt{package} goal of Maven's build lifecycle. It takes the ontologies that are part of the OntoMaven project and selects exactly those modules that are specified by the Maven parameter \texttt{userAspects}.
An additional parameter \texttt{aspectsIRI} allows to specify a custom OWL object or annotation property which is used to map the aspects to the ontology modules.
The parameter \texttt{ifIncludeOriginalAxioms} specifies whether those axioms in the ontology that are free of aspects (the \emph{base module}) should be included in the resulting ontology or not. This allows to either configure an ontology and enable selected aspects of it or to merely extract a module on its own and use it in the application.
An example configuration is shown in the following POM listing:

\begin{scriptsize}
\begin{verbatim}
<build> <plugins> <plugin>
    <groupId>de.csw.ontomaven</groupId>
    <artifactId>OntoMvnApplyAspects</artifactId>
    <version>1.0-SNAPSHOT</version>
    <configuration>
        <userAspects>
            <aspect>http://example.org/reputation#Reputation123</aspect>
            <aspect>http://example.org/provenance#prov_789</aspect>
        </userAspects>
        <aspectsIRI>http://corporate-semantic-web.de/aspectOWL#hasAspect</aspectsIRI>
        <includeOriginalAxioms>true</includeOriginalAxioms>
    </configuration>
...
\end{verbatim}
\end{scriptsize}

\subsection{Other Plug-Ins}
\label{sub:others}

The \texttt{OntoMvnSvn} and \texttt{OntoMvnGit} plug-ins provide ontology versioning support for OntoMaven. They use extensions to Subversion and Git, respectively. The Subversion extension is called SVont\footnote{\scriptsize{\url{http://www.corporate-semantic-web.de/svont.html}}} \cite{conf/gi/Luczak-RoschCPRT10}, which can compute semantic differences\footnote{\scriptsize{for the decription logic $\mathcal{EL}$}} for the versioning of ontologies. SVont supports typical Subversion commands such as \texttt{checkout}, \texttt{status}, \texttt{diff}, \texttt{commit}, and \texttt{info}.

The implementation of the commands \texttt{status} and \texttt{diff} and the plug-in \texttt{OntoMvnReport} are described in more detail in \cite{paschke:2013aa}.

\begin{figure}
  \centering
  \includegraphics[width=2cm]{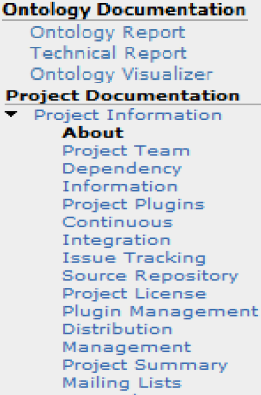}
  \includegraphics[width=6cm]{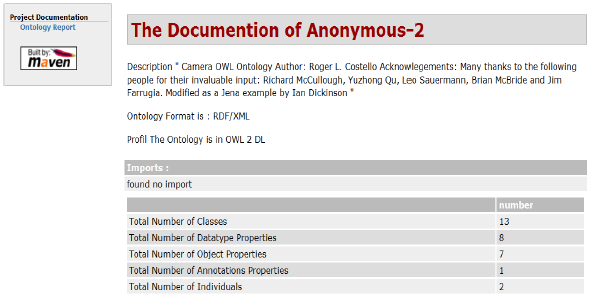}
   \includegraphics[width=6cm]{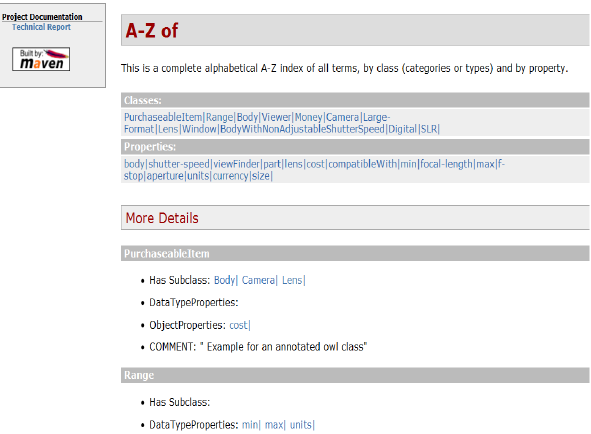}
  \caption{OntoMaven Summary and Report Views}\label{fig:ProjectDocumentation}
\end{figure}

The \emph{ontology report summary} is created by the goal \texttt{ontologyreport}. The middle figure \ref{fig:ProjectDocumentation} shows an example ontology summarization which gives an overview about the general description, the format, the semantic profile, imported ontologies and a summary about the ontology's statistics (number of classes, datatype properties, object properties, etc.).

For the documentation of the ontology, the plug-in uses existing automated ontology documentation tools. We have integrated the SpecGen ontology documentation tool which creates a HTML page containing detailed information about the classes and the properties. We further extended SpecGen with various algorithms for creating structure based concept groupings. \cite{Coskun2012,DBLP:conf/womo/CoskunRTP11} These groupings are used as basis for a visual documentation of the ontology. To support this process of creating such concept groups for the documentation of ontologies we extended the SpecGen tool with an automatic concept grouping functionality and embedded it for the OntoMaven documentation.

More detailed reports in the form of \emph{technical ontology report} and \emph{network visualizations} may be created by the goals \texttt{technicalreport} and \texttt{visualizer}, respecively (see right side of Figure \ref{fig:ProjectDocumentation}). The latter uses different graph visualizations \footnote{\scriptsize{\url{http://www.corporate-semantic-web.de/ontology-modularization-framework.html}}}.

%

\subsection{OntoMvnTest}

The \emph{OntoMvnTest} plug-in implements functionalities for the test phase. The plug-in executes the configured tests using the goal \texttt{test}. It is also used internally in other phases such as the \texttt{package} goal. The plug-in implementation uses the Pellet reasoner\footnote{\scriptsize{\url{http://clarkparsia.com/pellet/}}} to execute the ontology test cases. As default test suites the plug-in supports the W3C OWL Test Cases. This test collection contains different types of test cases, such as a test that determines and returns the OWL sublanguage, tests for inconsistency checks, and entailment tests, which test if the intended conclusions (represented by an output ontology) are entailed in the input ontology model. For instance, the intended entailment test result values are \texttt{Entailment} (positive test result) or \texttt{NoEntailment} (negative test result).

%

\section{Related Work Evaluation}
\label{relatedwork}

There are many existing ontology engineering methodologies and ontology editors available. The focus of e.g. Protege\footnote{\scriptsize{\url{http://protege.stanford.edu/}}}, Swoop\footnote{\scriptsize{\url{http://www.mindswap.org/2004/SWOOP/}}}, and Top Braid Composer\footnote{\scriptsize{\url{http://www.topquadrant.com/products/TB_Composer.html}}} is in the front-end providing user interfaces for ontology modeling / representation. Other editors support, e.g., visual ontology modeling such as Thematix Visual Ontology Modeler (VOM)\footnote{\scriptsize{\url{http://thematix.com/tools/vom/}}}, which enables UML-based ontology modeling based on the OMG Ontology Definition Metamodel (OMG ODM\footnote{\scriptsize{\url{http://www.omg.org/spec/ODM/}}}), or lightweight domain specific vocabulary development, such as Leone\footnote{\scriptsize{leone - \url{http://www.corporate-semantic-web.de/leone.html}}} \cite{PA155}. OntoMaven mainly differs from them in the Maven-based approach how it manages and declaratively describes the development phases, goals, and artifacts in a Maven Project Object Model (POM). In particular, OntoMaven has a focus on managing and (re-) using existing distributed ontologies in the implementation of ontology-based applications, while the editors have a focus on supporting the modelling of ontologies for later use. That is, OntoMaven is not a full ontology modeling tool as e.g., Protege, Swoop, and Top Braid Composer, which provide a development user interface. Instead OntoMaven's has its strength in the back-end management of distributed ontology modules including support for reuse (transitive imports), dependency management and collaboration (semantic versioning). Further implementation-specific and plug-in-specific differences are in the underlying details of the provided functionalities of OntoMaven such as POM-based dependency management, semantic versioning, semantic documentation etc. For a comparison see Table \ref{ToolComparison}.

\begin{table}
\caption{Functional Comparison of OntoMaven with Ontology Development Tools} \label{ToolComparison}
\centering
\begin{scriptsize}
\begin{tabular}{|p{2cm}|p{3cm}|p{3cm}|p{3cm}|p{3cm}|}
  \hline
   & OntoMaven & Protege & Swoop & Top Braid Composer \\
  \hline
  Repositories & yes (local and remote) & yes (local and remote) & no & yes (by Allegro Graph 4 PlugIn) \\
  \hline
  Reuse (Import) & yes (dependency management) & yes & yes & yes \\
  \hline
  Collaboration (Versioning) & yes (semantic diff) & no & no & no \\
  \hline
  Documentation & yes (text and visual) & yes (text and visual) & yes (only text) & yes (text and visual in Maestro version) \\
  \hline
  Testing & yes & yes & yes & yes \\
  \hline
  Extensibility & yes & yes (many existing plugins) & yes & yes (commercial) \\
  \hline
\end{tabular}
\end{scriptsize}
\end{table}

The W3C Wiki lists several existing ontology repositories\footnote{\scriptsize{\url{http://www.w3.org/wiki/Ontology_repositories}}}. Further ontology repositories are, e.g., COLORE\footnote{\scriptsize{\url{http://stl.mie.utoronto.ca/colore/}}} for ontologies written in the ISO Common Logic (CL) ontology languages and Ontohub\footnote{\scriptsize{\url{http://ontohub.org/}}} which maintains a set of heterogenous ontologies. The current focus of these projects is on collecting and listing of existing ontologies. Apart from simple search functionalities there is no support for repository-based ontology development which is the focus of OntoMaven and OntoMaven repositories.

New standardization efforts such as OMG Application Programming Interfaces for Knowledge Bases (OMG API4KB)\footnote{\scriptsize{\url{www.omgwiki.org/API4KB/}}} aim at the accessability and interoperability of heterogenous ontologies via standardized programming interfaces. OntoMaven can act as one possible implementation of the development support functionalities of OMG's API4KB.  With its Maven-based approach for structuring the development phases into different goals providing different functionalities during the software development project's life cycle, OntoMaven supports in particular agile ontology development methods, such as COLM \cite{RoeHe09}, as well as development methods which are inherently based on modularization such as aspect-oriented ontology development \cite{SchPa13}.

\section{Conclusion}
\label{conclusion}

Apache Maven is a widespread and highly successful tool in Software Engineering for build automation and development project life cycle management. The contribution in this paper is a new design artifact called OntoMaven which adapts Maven for ontology-based development and dynamic configuration and use of ontology modules by the means of Aspect-oriented ontologies (AOOD). It is built using Maven's plugin-based architecture. The proof-of-concept of OntoMaven implements several useful plugins which interface with existing ontology development tools and functionalities such as the plug-ins \emph{OntoMvnImport}, \emph{OntoMvnApplyAspects}, \emph{OntoMvnSVN}, \emph{OntoMvnReport}, and \emph{OntoMvnTest}.

\section{Acknowledgements}
This work has been partially supported by the InnoProfile Transfer project "Corporate Smart Content" funded by the German Federal Ministry of Education and Research (BMBF).

\bibliographystyle{plain}
\bibliography{bibliography}

\end{document}